# Interaction phenomena in graphene seen through quantum capacitance

G. L. Yu[a], R. Jalil[b], Branson Belle[b], Alexander S. Mayorov[a], Peter Blake[b], Frederick Schedin[b], Sergey V. Morozov[c], Leonid A. Ponomarenko[a], F. Chiappini[d], S. Wiedmann[d], Uli Zeitler[d], Mikhail I. Katsnelson[e], A. K. Geim[a,b], Kostya S. Novoselov[a], and Daniel C. Elias[a]

[a]School of Physics and Astronomy, University of Manchester, Manchester, M13 9PL, United Kingdom;

[b]Centre for Mesoscience and Nanotechnology, University of Manchester, Manchester, M13 9PL, United Kingdom;

[c]Institute for Microelectronics Technology, 142432 Chernogolovka, Russia;

[d]High Field Magnet Laboratory, Radboud University Nijmegen, NL-6525 ED Nijmegen, The Netherlands;

[e]Institute for Molecules and Materials, Radboud University Nijmegen, NL-6525 AJ Nijmegen, The Netherlands

*Capacitance measurements provide a powerful means of probing the density of states. The technique has proved particularly successful in studying 2D electron systems, revealing a number of interesting many-body effects. Here, we use large-area high-quality graphene capacitors to study behavior of the density of states in this material in zero and high magnetic fields. Clear renormalization of the linear spectrum due to electron–electron interactions is observed in zero field. Quantizing fields lead to splitting of the spin- and valley-degenerate Landau levels into quartets separated by interaction-enhanced energy gaps. These many-body states exhibit negative compressibility but the compressibility returns to positive in ultrahigh B. The reentrant behavior is attributed to a competition between field-enhanced interactions and nascent fractional states.*

The Dirac-like spectrum of charge carriers in graphene (1) gives rise to a constant ratio between their kinetic and Coulomb energies (2). The ratio is given by the coupling constant $\alpha = e^2/\varepsilon\hbar v_F$, where $e$ is the electron charge, $\hbar$ is the reduced Planck constant, and $\varepsilon$ is the effective dielectric constant (2). Because the Fermi velocity $v_F$ is 300× smaller than the speed of light $c$, $\alpha$ is close to unity, that is, much larger than the fine-structure constant $e^2/\hbar c$. This regime of strong relativistic-like coupling presents considerable interest from the point of view of many-body physics (2). For example, the large $\alpha$ leads to a noticeable renormalization of $v_F$ in the vicinity of the Dirac point (3), which has clear analogies with quantum-field theory (4). In the presence of a magnetic field $B$, many-body effects become more pronounced, generating large interaction-enhanced gaps at filling factors $\nu = 0, \pm 1, \pm 3, \pm 4, \pm 5, \pm 7, ...$ (5). Still, despite the intensive research in recent years (2), many-body physics in graphene is far from complete.

Unlike in the conventional 2D systems, the density of states (DoS) in graphene depends on $n$; in capacitance measurements, this makes even small DoS contributions readily noticeable on top of a constant geometrical capacitance $C_G$. This allowed several recent observations of graphene's quantum compressibility (6–12). However, considerable charge inhomogeneity typical for graphene deposited on silicon oxide leads to strong spatial averaging. For example, graphene-on-$SiO_2$ devices normally exhibit pronounced quantum Hall effect (QHE) features, but Landau quantization in their total capacitance $C$ is seen only as weak oscillations (9). This inhomogeneity obscures finer details in the DoS which can indicate new phenomena.

In this article, we use graphene deposited on hexagonal boron nitride (hBN), which has dramatically reduced charge inhomogeneity. Our devices have relatively large area, $S \sim 10^3$ μm$^2$, to increase their capacitance $C$ but, despite this, quantum oscillations are pronounced already in $B$ below 1 T and correspond to changes in $C$ by a factor of >10. The device quality has allowed accurate measurements of their DoS which in zero $B$ reveal the same renormalization behavior of $v_F$ as previously reported (3) in quantizing $B$, where the existing renormalization theory is, strictly speaking, inapplicable (2). In finite $B$, we observe the interaction-induced breaking of the fourfold degeneracy of zero Landau level (LL) with gaps at filling factors 0 and ±1 being of the same order of magnitude, in agreement with the SU(4) isospin symmetry suggested for graphene (13). Furthermore, our capacitors are found to exhibit extended regions of negative compressibility, which can be associated with quenching of Dirac fermions' kinetic energy by $B$. These regions disappear in $B > 25$ T, probably due to the development of a spatially inhomogeneous fractional QHE state such that their contribution to the compressibility is on average positive.

**Measurements and Devices**
Differential capacitance $C$ can be described (14–19) in terms of $C_G$ and the quantum capacitance $C_Q$ which act in series

$1/C = 1/C_G + 1/C_Q;$ [1]

where

$C_Q = Se^2 dn/d\mu.$ [2]

Here, $S$ is the area of the electrodes, $n$ is the carrier concentration, and $\mu$ is the chemical potential, so that $dn/d\mu$ is the thermodynamic DoS. To understand the origin of the quantum capacitance term, consider the case of a parallel plate capacitor with one electrode made of a 2D metal and the other made of a normal metal having a large DoS (Fig. 1). As seen from Fig. 1c, the difference in electrochemical potentials $V$ is determined by the electric potential drop $\varphi$ between the two electrodes and the shift in $\mu$

$eV = e\varphi + \mu.$ [3]

By differentiating 3, one immediately arrives at 1.

Our experimental devices were composed of graphene and a Ti(5 nm)/Au(50 nm) film as two electrodes separated by a hBN crystal, typically 20–30 nm thick. The graphene layer rested on another hBN layer to improve the graphene's electronic quality (20, 21) (Fig. 1a). The sandwich structure was prepared by using the dry transfer method (22) on top of a quartz substrate. The latter was essential to minimize the parasitic capacitance that otherwise did not allow the use of the standard oxidized Si wafers as a substrate. Transfer of large graphene flakes on flat surfaces resulted in many bubbles filled with either air or hydrocarbon residue (23, 24). By using electron-beam lithography, we designed our top electrodes so that they did not cover bubbles. This proved to be important to achieve high charge homogeneity over the whole device area.

The differential capacitance was measured by using a capacitance bridge at a 1 kHz excitation frequency. The excitation amplitude was in the range 5–25 mV, carefully chosen for each device, so that the induced oscillations in μ were below the fluctuations caused by charge inhomogeneity.

**Zero-Field DoS.** An example of our capacitance measurements as a function of bias $V$ applied between the two electrodes is presented in Fig. 2a. $C$ exhibits a sharp minimum near zero bias and tends to saturate at large positive and negative $V$. We attribute the behavior to a small value of $C_Q$ associated with the low DoS near the neutrality point (6–12). Indeed, the DoS in graphene is given by

$$dn/d\mu = 8|\mu|\pi/h^2 v_F^2 ; \qquad [4]$$

where $h$ is the Planck constant. Near $\mu \approx 0$, $C_Q$ approaches zero, and the second term in 1 dominates the measured capacitance. With increasing bias, μ moves toward higher DoS in graphene's energy spectrum and the effect of $C_Q$ decreases.

By combining 1, 2, and 4 and using the expression $C_G = \varepsilon_0 \varepsilon_{BN} S/d$ for our parallel-plate capacitors, we can fit the measured curves with essentially only one parameter $v_F$. To this end, thickness $d$ of hBN was found by using atomic force microscopy, and hBN's dielectric constant $\varepsilon_{BN} \approx 4.5$ is known from literature. In addition, the ratio $d/\varepsilon_{BN}$ that comes into the final expression for $C$ is found independently from measurements of quantum oscillations in finite $B$ (see below). Thus, noticing that the device area $S$ was determined by optical and scanning electron microscopy within 5% accuracy, $C_G$ can be determined directly, without any fitting parameter. The fit, shown in Fig. 2a, yields $v_F^0 = 1.05 \times 10^6$ m/s for all our devices, in excellent agreement with the values reported in transport experiments (25, 26). Note that a small parasitic capacitance (associated with wiring, usually on the order of a few tens of femto-Farad, depending on the particular setup used) has been subtracted in such a way that at a very high carrier concentration the measured capacitance is equal to $C_G$ (which is known without any fitting parameter).

Although the experimental and theoretical curves in Fig. 2a practically coincide, the high accuracy of our measurements allows further comparison between experiment and theory by zooming in at minute differences between the curves, which are not visible on the scale of Fig. 2a. To zoom in, we can replot our data in terms of $v_F$. By using the standard expression $\mu = (n/\pi)^{1/2} h v_F(n)/2$ to substitute μ in 3, we arrive at $v_F(n) = 2(\pi/n)^{1/2}(eV - ne^2 d/\varepsilon_0 \varepsilon_{BN})/h$. Fig. 2b shows the experimental dependence $v_F(n)$, where $n$ was obtained self-consistently by integrating the non-constant differential capacitance over the corresponding range of $V$ (note that the standard approximation $n \propto V$ fails for strongly varying $C$). One can see that $v_F$ significantly depends on $n$ but lies around the average value of $v_F$ reported above.

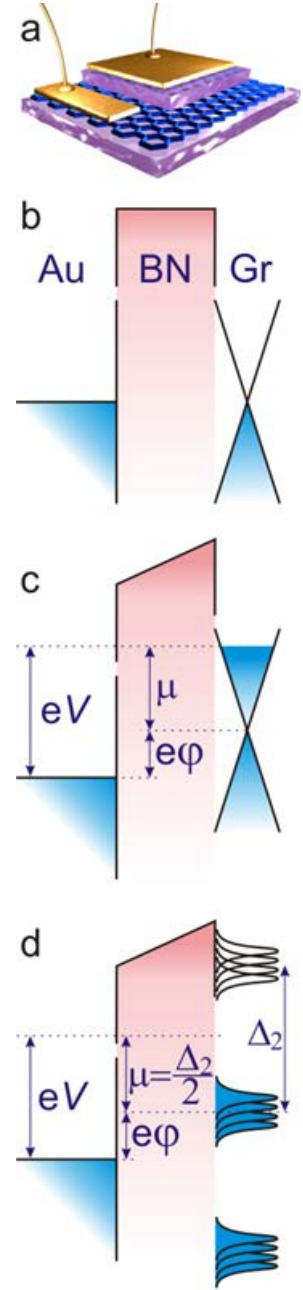

Fig. 1. Schematics of our experiments. (a) Graphene capacitors. Graphene crystal (blue hexagonal layer) is contacted by a gold pad. Second electrode is deposited onto thin hBN (purple). (b) Schematic band diagram at zero bias. (c) Same as (b) but at a finite bias. (d) Band diagram at finite $B$ with the Fermi energy μ in graphene at a filling factor ν = 2. Each LL is shown to be 4× degenerate.

To explain the observed deviations from a constant value of $v_F$, we recall that electron–electron interactions lead to spectral renormalization. Many-body contributions to the compressibility of graphene were calculated first from the total energy (27) and turned out to be logarithmically divergent at the neutrality point. Diagrammatically, there are two types of many-body effects on the compressibility, namely, renormalization of the single-particle DoS (that is, of the Fermi velocity) and vertex corrections. For massless Dirac fermions, the latter does not contain any big logarithms; thus, with the logarithmic accuracy, the renormalization of the compressibility is determined only by the renormalization of the Fermi velocity, such that $v_F$ logarithmically diverges near the Dirac point (2, 28, 29). The renormalization can be described by (3)

$$v_F(n) = v_F(n_0)[1+e^2 \cdot \ln(n_0/n)/16h\varepsilon_0\varepsilon v_F(n_0)]. \quad [5]$$

The theory fit shown in Fig. 2b is obtained by using $\varepsilon = 8$, cutoff density $n_0 = 10^{15}$ cm$^{-2}$, and single-particle Fermi velocity for graphene $v_F(n_0) = 0.85 \times 10^6$ m/s. The latter two parameters are taken from theory and the earlier reports (2, 3, 30). For graphene on hBN, $\varepsilon$ has to take into account both dielectric screening and self-screening in graphene, and the random-phase approximation (2) suggests $\varepsilon = \varepsilon_{BN} + \pi e^2/4hv_F\varepsilon_0$, which yields $\varepsilon \approx 7.8$ for our average $v_F = 1.05 \times 10^6$ m/s. The agreement between experiment and theory in Fig. 2b might be better than that reported for suspended graphene (3), even though changes in $v_F$ are smaller in the present case. The smallness is because of extra screening by hBN and higher charge inhomogeneity (22, 31) that limits the minimum achievable $n$ to $\sim 10^{10}$ cm$^{-2}$. We emphasize that, unlike ref. 3, we do not use quantizing $B$ to deduce $v_F(n)$, in which case the use of the zero-$B$ renormalization theory has not been justified.

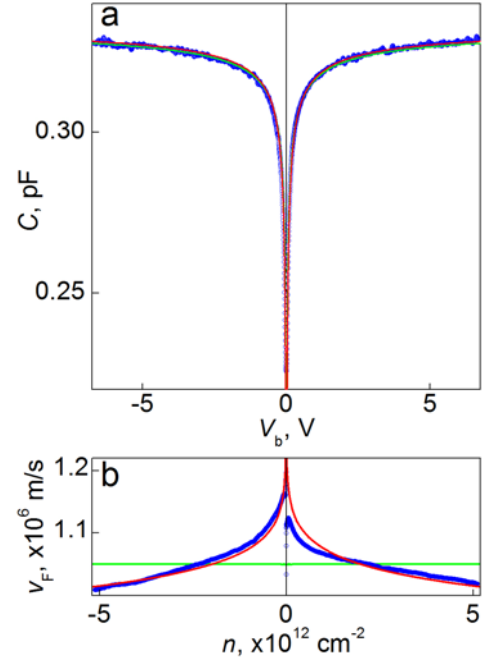

Fig. 2. Quantum capacitance of graphene. (a) Differential capacitance in zero $B$. Blue symbols are experimental data; green and red curves are the best fits with constant and renormalized $v_F$, respectively. This particular device has $d \sim 27$ nm and $S \sim 250$ μm$^2$. (b) Same data replotted in terms of $v_F$ and carrier concentration $n$; color coding as in (a).

**Quantum Capacitance and Gaps in High Fields.** Fig. 3 summarizes results of our capacitance measurements in quantizing $B$. At relatively small $B < 3$ T a series of sharp minima is observed as function of $V$, which reveals minima in the DoS due to large cyclotron gaps between the fourfold-, spin-, and valley-degenerate LLs (Fig. 3a). In the first approximation, the periodicity $\Delta V$ of the capacitance minima is given by the amount of charge required to fill in each LL, which leads to the equation $C_G\Delta V = 4Se^2B/h$. As noted above, this provides a convenient way to determine the effective thickness of hBN, $d/\varepsilon_{BN} = h\Delta V/(4e^2B)$. In $B > 3$ T the degeneracy of zero LL gets lifted, which results in three additional minima at $\nu = 0$ and $\pm 1$. In even higher $B > 10$ T, the other LLs also start splitting into quartets (Fig. 3b and c). Note that it is usually difficult for transport measurements to quantify the gap at the neutrality point (5, 13, 32). Our capacitance studies show that in terms of the DoS there is no principal difference between this gap and those at $\nu = \pm 1$, and all of them exhibit similar behavior (Fig. 3).

We have used three different methods to evaluate the energy gaps $\Delta_\nu$ for different $\nu$. The first one is the standard fitting of thermal smearing of quantum oscillations (minima in $C$) with the Lifshitz–Kosevich formula (3, 9, 25, 26). The second approach is based on calculating $\mu(n)$ by integrating the reciprocal quantum capacitance $1/C_Q \propto d\mu/dn$. The results of such integration are shown in Fig. 4a, where the steps in $\mu$ around the integer $\nu$ indicate the corresponding gaps. The third method is to monitor positions of the capacitance minima as a function of $B$ and analyze deviations from the linear behavior that is expected when the effect of quantum capacitance is negligibly small. The latter method uses Eq. 2 and the fact that the $\nu = 2$ minima should occur at $\mu = v_F(ehB/4\pi)^{1/2}$, half the cyclotron gap (Fig. 1e). We note that at least the first and third methods do not suffer from the usual problem of the capacitance measurements—incomplete charging of the sample due to its large resistance (the first method relies on high-temperature data, where resistance is small, and the third method only considers the position of the minima of the capacitance, not its value).

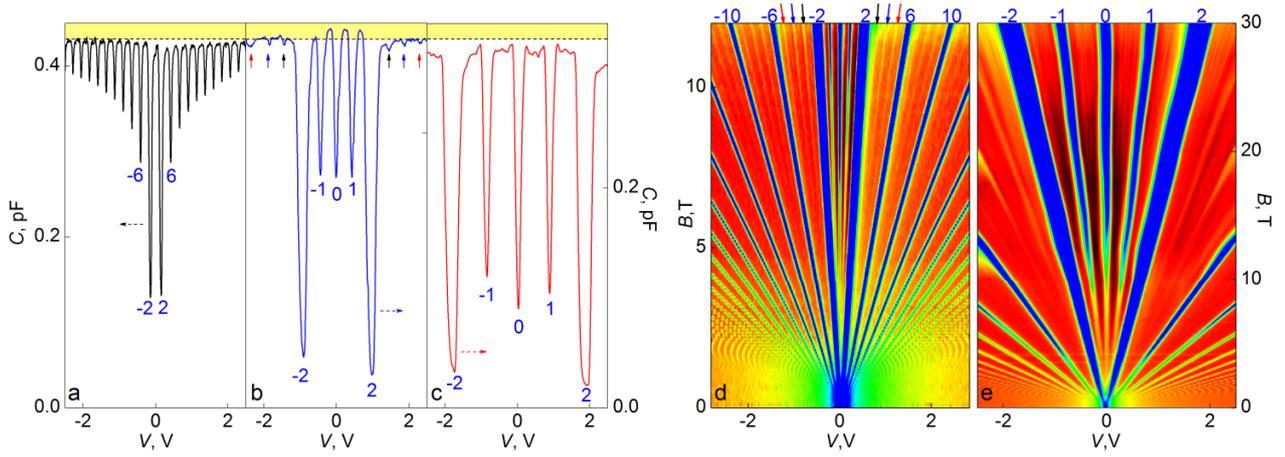

Fig. 3. Graphene capacitors in quantizing fields. (a–c) Examples of graphene capacitance for $B$ = 3, 15, and 30 T, respectively. Blue numbers are $\nu$ for the corresponding minima. The arrows in b mark $\nu = \pm 3$ (black arrows), $\pm 4$ (blue) and $\pm 5$ (red). The dashed line indicates $C_G = 0.433 \pm 0.002$ pF (for a device shown in a) and $0.335 \pm 0.002$ pF (for a device shown in b and c). The 15 T curve reveals that the total capacitance becomes higher than the geometrical one at $\nu$ around $\pm 1/2$ and $\pm 3/2$, indicating a negative contribution from $C_Q$, that is, negative compressibility. (d) Two-dimensional map of differential capacitance as a function of $B$ and $V$ for the same device as in a. Color scale is 0.37 pF to $C_G$ (blue to green to red) and $C_G$ to 0.454 pF (red to black). Numbers and arrows are as in b. (e) Two-dimensional map in $B$ up to 30 T for the device in b and c. Scale: 0.23 pF to $C_G$ (blue to green to red) and $C_G$ to 0.349 pF (red to black). The dark regions (d and e) correspond to negative $C_Q$.

The results of our measurements for $\Delta_\nu$ are shown in Fig. 4b. For the main gap, all three methods yield essentially the same size and behavior of $\Delta_{\pm 2}$ and agree well with graphene's single-particle description (1). Nonetheless, we point out that in $B < 3$T there is a systematic tendency for the measured values to be higher than the expected single-particle gap (Fig. 4b), and a better agreement can be achieved by using the renormalized $v_F$ to fit $\Delta_{\pm 2}(B)$.

The situation with the many-body gaps at $\nu = 0$ and $\pm 1$ is more complicated. The temperature dependence and the integration of $1/C_Q$ (first and second methods) yield similar absolute values for $\Delta_\nu$ in high $B$ (Fig. 4b) but qualitatively different functional behavior on $B$ (square-root and superlinear, respectively). The third approach is inapplicable in this case as it cannot distinguish between contributions coming from $\Delta_0$ and $\Delta_{\pm 1}$. Both gaps are 1 order of magnitude larger than what would be expected from Zeeman splitting with g-factor equal to 2, in agreement with transport measurements (5, 13). The same size and behavior of the many-body gaps do not allow us to distinguish between spin and valley polarizations for gaps at $\nu = 0$ and $\pm 1$, which might suggest nontrivial broken-symmetry states in accordance with the SU(4) model. We have carried out additional experiments in tilted $B$ to probe which of the states exhibits a stronger spin contribution but found no noticeable difference. As for the different dependences $\Delta_\nu(B)$ observed by the two methods, this remains to be understood but is not totally unusual because many-body contributions can contribute differently in different observables. Also, the fact that the many-body gap might depend on the temperature can influence the obtained value of the gap when using the Lifshitz–Kosevich formula (the first method).

**Negative Compressibility.** We turn our attention to the capacitance behavior around half-integer $\nu$. Figs. 3 and 4 show that for $B$ between 5 and 25 T, the total capacitance can exceed the geometrical one, indicating negative $C_Q$. The regime of negative compressibility is seen clearly on the 2D maps of Fig. 3 as the dark areas around $\nu = \pm 1/2$ and $\pm 3/2$. The yellow shaded areas in Figs. 3 and 4 mark this regime, too. Weaker signatures of negative $C_Q$ were also observed around $\nu = \pm 5/2, \pm 7/2, \pm 9/2$, and $\pm 11/2$ (e.g., see the dark areas in Fig. 3e). In principle, it is well known that interaction effects can result in negative compressibility (16, 18, 19). In the conventional 2D systems, the kinetic energy scales as $n$ whereas the potential energy scales as $n^{1/2}$; therefore, the negative compressibility regime can be reached at sufficiently low $n$. In graphene, both energies scale as $n^{1/2}$ and negative $dn/d\mu$ is not expected for any $n$. However, if the kinetic energy is quenched by quantizing $B$, this leaves only Coulomb interactions and can lead (17, 33) to negative $dn/d\mu$ for small values of $d/l_b$, in agreement with the observed behavior (for our devices, $d \sim l_b$ in 1 T). It is expected (33) that as $B$ increases and $d/l_b$ becomes large, the compressibility at half-integer $\nu$ should tend to zero and therefore $C$ saturates to its geometrical value.

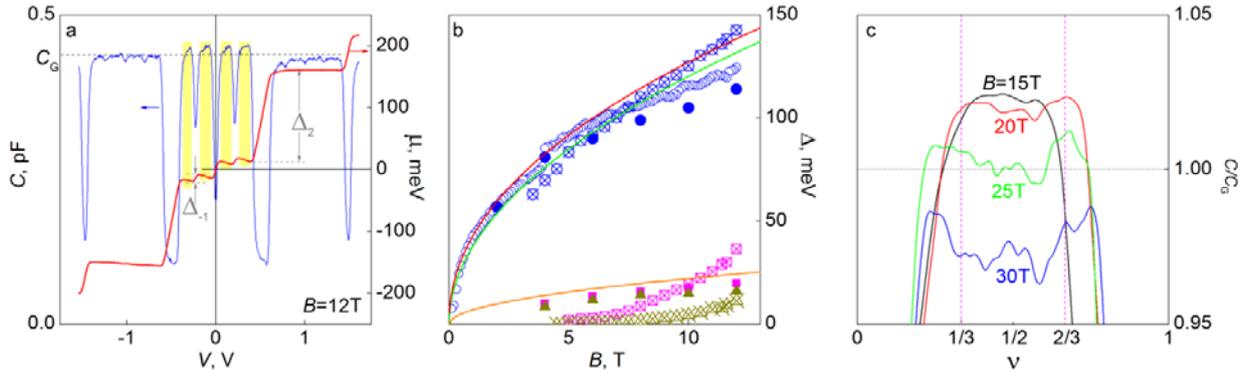

Fig. 4. Many-body gaps through capacitance measurements. (a) $C(V)$ for one of our devices at $B = 12$ T and 3 K (blue curve). Red curve: the corresponding changes in $\mu$. Yellow shaded areas mark the regions of negative compressibility. (b) Energy gaps found for different $\nu$. Symbols correspond to alternative methods of determining $\Delta$. Solid symbols: by using temperature dependence of the capacitance minima; crossed symbols: from steps in $\mu$ as shown in a; open symbols: from magnetic field dependence of the minima's positions. Blue circles: $\nu = 2$ (similar results for $\nu = -2$); magenta squares: $\nu = 0$; olive triangles: $\nu = \pm 1$. Green curve: expected cyclotron gap for $v_F = 1.05 \times 10^6$ m/s; red curve: same but using the renormalized values of $v_F(n)$; orange: Coulomb energy $e^2/4\pi\varepsilon l_b$, where $l_b = (h/2\pi eB)^{1/2}$ is the magnetic length. (c) Recovery of quantum capacitance from negative to positive with increasing high $B$. For this device, $d = 28$ nm; 2 K. Black curve corresponds to $d/l_b \approx 3.8$; red: 4.5; green: 5.1; blue: 5.9.

The latter prediction is in conflict with our experiment that shows a reentrant behavior such that in high $B$, negative $C_Q$ decreases but then becomes positive again above 25 T (Fig. 3e). This reentry is shown in more detail in Fig. 4c for $\nu = 1/2$. We attribute the behavior to an additional positive contribution coming from nascent fractional QHE states (12, 18, 19, 31, 34) that are not individually resolved because of charge inhomogeneity but, nevertheless, appear locally as reported in refs. (12) and (34). Due to spatial averaging of these fractional QHE contributions, a positive average $C_Q$ can be expected in high $B$ near $\nu$ that show strongest fractional states such as 1/3 and 2/3 (Fig. 4c).

**Acknowledgements**.

We are grateful to B. Skinner, B. I. Shklovskii, and L.Eaves for useful discussions. This work was supported by the European Research Council, European Commission Seventh Framework Programme, Engineering and Physical Research Council (UK), the Royal Society, US Office of Naval Research, US Air Force Office of Scientific Research, US Army Research Office, the Körber Foundation, and EuroMagNET II under the European Union Contract 228043.

**References.**